\documentclass[aps,prl,11pt,twoside,nofootinbib]{revtex4-2}

\usepackage{amsmath,latexsym,amssymb,verbatim,enumerate,graphicx}

\usepackage{float}
\usepackage[english]{babel}
\usepackage[T1]{fontenc}
\usepackage{appendix}
\usepackage{tikz}
\usepackage{braket}
\usepackage{anyfontsize}
\usepackage[11pt]{moresize}
\usepackage{enumerate}
\usepackage{dcolumn}
\usepackage{bm}
\usepackage{tabularx}
\usepackage{array}
\usepackage{lipsum}
\usepackage{cancel}
\usepackage{enumitem}
\usepackage{quantikz}
\usepackage{graphicx, caption, subcaption,overpic,float}

\usepackage{theorem}

\usepackage{color}

\newcommand{\be}{\begin{equation}}
\newcommand{\ee}{\end{equation}}
\newcommand{\bea}{\begin{eqnarray}}
\newcommand{\eea}{\end{eqnarray}}

\providecommand{\U}[1]{\protect\rule{.1in}{.1in}}

\def\1{\openone}

\begin{document}

\title{Bayesian and Markovian classical feedforward \\ for discriminating { qubit} channels}

\author{Milajiguli Rexiti}
 \email{rexiti.milajiguli@unicam.it}
\author{Stefano Mancini}
 \email{stefano.mancini@unicam.it}
 \affiliation{School of Science and Technology, University of Camerino, Via Madonna delle Carceri 9,62032 Camerino, Italy\\
    INFN, Sezione di Perugia, Via A. Pascoli, 06123 Perugia, Italy}

\begin{abstract}
We address the issue of multishot discrimination between { two qubit channels} by invoking a simple adaptive protocol that employs Helstrom measurement at each step and classical information feedforward, { beside separable inputs.}
We contrast the performance of Bayesian and Markovian strategies. We show that the former is only slightly advantageous and for a limited parameters' region.
\end{abstract}

\maketitle


\section{Introduction}

The notion of a quantum channel has become ubiquitous in quantum information
processing. Thus, the ability to discriminate between two (or more) quantum channels has emerged as a { challenging} issue \cite{S05,WY06,RFY09}. 
Although more daunting, { the problem of quantum channel discrimination can be traced back to the task of distinguishing quantum states \cite{BK15}}. Thus, in the single-shot case, the ultimate limits can be established with the help of the Helstrom bound \cite{Hel69}.
When moving to the multi-shot scenario, if unconstrained input and measurement are allowed, the problem becomes similar to the single-shot case, although in a larger space. However, { this approach assumes access to costly resources such as entangled input states and non-local measurements}. For practical purposes, it would be worth considering the limits imposed by local resources, such as separable input states and local measurements.
In such a context, adaptive strategies { become relevant}. { Remarkably, certain cases have demonstrated that adaptive strategies can outperform non-adaptive ones (possibly involving entangled inputs and collective measurements)} \cite{HHLW10}. Thus, some efforts have been recently devoted to clarifying in which situations adaptive strategies offer or do not offer an improvement \cite{H09,BHKW20,PPKKO21,SHW22}. { Finding optimal adaptive strategy is known to be a semi-definite programming problem, which however scales exponentially with the number of channel uses \cite{Kat21}.}

{
Here, because of the higher complexity of adaptive methods, we are motivated to clarify, within adaptive protocols that feedforward at each step only classical information, the difference between Bayesian and Markovian strategies. The former exploits information gathered through the entire history, while the latter only that of the previous step. This will be done for some paradigmatic examples of qubit channels, namely the depolarizing channel, the bit-flip channel and the amplitude damping channel. 
For single shot, the discrimination within each of these classes of channels has been studied respectively in Refs.\cite{S05}, \cite{RMM22} and \cite{RM21}. Actually, we shall address the issue of multi-shot discrimination between two channels (be either depolarizing, or bit-flip or amplitude damping) using separable inputs (without side entanglement), by resorting  
to Helstrom measurement at each step and classical information feedforward (see Fig.\ref{fig1} left). We then contrast the performance of Bayesian and Markovian schemes within this strategy. Our findings show that the former is only slightly advantageous and for a limited parameter region.}


\section{One-shot discrimination}\label{sec:oneshot}

{ Suppose we want to distinguish between two qubit channels
characterized by parameters $\eta_0$ and $\eta_1$ and occurring with probabilities $p_0$ and $p_1$ ($p_0+p_1=1$). 
Let $\rho_{i}$, $i=0,1$, be the two channels' output for a given input.}
Then, the problem of channel discrimination is turned into the discrimination between two (generally mixed) states, $\rho_{0}$ and $\rho_{1}$, occurring with probabilities $p_0$ and $p_1$.
To this end, it is known \cite{Hel69} that the optimal measurement is given by the projection onto the positive and negative subspaces of the operator
\begin{equation}\label{eq:Delta}
\Delta\equiv p_0\rho_0-p_1\rho_1.   
\end{equation} 
However, this does not account for all possible cases. In fact,
the eigenvalues of $\Delta$ do not always split into positive and negative.
Thus, { we start presenting} hereafter an exhaustive solution for the binary discrimination of qubit states.

Let $\lambda_0,\lambda_1 (\lambda_0\geq\lambda_1)$ and $|v_0\rangle, |v_1\rangle$ be the ordered eigenvalues and the corresponding orthonormalized eigenvectors of the operator 
$\Delta$ in \eqref{eq:Delta}. The necessary and sufficient conditions 
for an optimal POVM $\{\Pi_0,\Pi_1\}$ in the qubit case read \cite{Hel69}:
\be\label{syseq}
\begin{cases}
\Delta\Pi_0\geq0 \\
-\Delta\Pi_1\geq0 \\
\Pi_0\Delta\Pi_1= 0
\end{cases}.
\ee
It is easy to see that when $\lambda_0>0>\lambda_1$, the optimal POVM can be constructed as $\{\Pi_0=|v_0\rangle\langle v_0|,\Pi_1=|v_1\rangle\langle v_1|\}$. Nevertheless, when $\Delta$ has all positive or all negative eigenvalues, the system of equation \eqref{syseq} has no solution except $\{{\rm I}, \bold 0\}$. However, the latter is not always optimal. Hence, we have to treat separately the cases where $\Delta$ has all non-negative (non-positive) eigenvalues.

Notice that any rank 1 projector in qubit space can be written as
\be\label{pj}
\begin{pmatrix}
\cos^2{\theta} && e^{-i\phi}\sin\theta \cos\theta\\
e^{i\phi}\sin\theta \cos\theta && \sin^2{\theta}
\end{pmatrix},
\ee
with $\theta\in [0,\frac{\pi}{2}], \phi\in [0, 2\pi]$. Then, assuming $\Pi_0$ as \eqref{pj} in the basis $\{|v_0\rangle,|v_1\rangle\}$ (and consequently $\Pi_1=I-\Pi_0$), the problem of maximizing the success probability becomes:
\bea
\max_{\Pi_0}\left[p_0 {\rm Tr}\left(\rho_0 \Pi_0\right)+(1-p_0) {\rm Tr}\left(\rho_1 \Pi_1\right)\right]&=&\max_{\Pi_0}\left[{\rm Tr}\left(\Delta \Pi_0\right)+(1-p_0)\right]\notag\\
&=&\max_{\theta}\left[\lambda_0 \cos^2{\theta}+\lambda_1 \sin^2{\theta}\right] { +(1-p_0).}
\eea
After checking the first and second derivatives and comparing with the set of projectors $\{{\rm I}, \bold 0\}$, we get the following results:
\begin{subequations}\label{optPOVM}
\begin{align}
& 1) \; \text{if $\lambda_0>0$ and  $2p_0\leq 1+\lambda_0$, we have $P_{succ}=\lambda_0+1-p_0$, being $\Pi_0=|v_0\rangle\langle v_0|$;}\\
& 2) \; \text{if $\lambda_0>0$ and $2p_0 > 1+\lambda_0$, we have $P_{succ}=p_0$, 
being $\Pi_0=I$;}\\
& 3) \;\text{if $\lambda_0<0 $ and $p_0\leq\frac{1}{2}$, we have $P_{succ}=1-p_0$, 
being $\Pi_0=0$;}\\
& 4) \; \text{if $\lambda_0<0 $ and $p_0>\frac{1}{2}$, we have $P_{succ}=p_0$, 
being $\Pi_0=I$.}
\end{align}
\end{subequations}
Notice that the condition $\lambda_0>0>\lambda_1$ is included in case 1). 

The probability of the measurement outcome $i$ ($i=0,1$), 
given the state $\rho_i$, reads
\begin{subequations}\label{contributionsPs}
\begin{align}
P_{\rho_0}(0)&\equiv {\rm Tr}\left(\rho_0 \Pi_0\right),\\
P_{\rho_1}(1)&\equiv {\rm Tr} \left(\rho_1\Pi_1\right).
\end{align}
\end{subequations}
Assuming that $\rho_{0}$, $\rho_{1}$ occur each with probability $1/2$,
the probability of success in discriminating between them follows as 
\be\label{Psuccess}
P_{succ}=\frac{1}{2}P_{\rho_0}(0)+\frac{1}{2}P_{\rho_1}(1).
\ee

{ 
Below we shall analyze illuminating examples of qubit channels and assume w.l.o.g.
an input state $\psi=|\psi\rangle\langle\psi|$ with
\begin{equation}\label{ketpsi}
\ket\psi=\sqrt{1-r}\ket{0}+e^{-i\varphi}\sqrt{r}\ket{1},
\end{equation}
where $ r\in[0,1]$ and $ \varphi\in[0, 2\pi)$, being $\{\ket{0},\ket{1}\}$ the canonical basis of $\mathbb{C}^2$.
We will also assume w.l.o.g. $\eta_0>\eta_1$.
}


{

\subsection{Depolarizing channel}

The qubit depolarizing channel is described by the map \cite{MWbook}
\begin{equation}
\psi\mapsto {\cal D}_\eta(\psi)=(1-\eta)\psi+\eta I/2,
\end{equation}
being $\eta\in[0,1]$.

One can easily realize that Eq.\eqref{Psuccess} gives
\begin{equation}
P_{succ}=\frac{1}{2}\left[1+\frac{1}{2} \left(\eta_0-\eta_1\right) \right],
\end{equation}
which is independent of the input state, that is, from parameters $r$ and $\varphi$.


\subsection{Bit-flip channel}

The (qu)bit-flip channel is described by the map \cite{MWbook}
\begin{equation}
\psi\mapsto{\cal F}_\eta(\psi)=(1-\eta)\psi+\eta X\psi X,
\end{equation}
being $\eta\in[0,1]$ and $X=|0\rangle\langle 1|+|1\rangle\langle 0|$ the bit-flip operator in the canonical 
basis of $\mathbb{C}^2$.

Since the action of the channel is the same in the $y$ and $z$ directions of the Bloch sphere, 
for the input state w.l.o.g. we can confine our attention to the $x-z$ plane. 
Hence, we choose $\varphi=0$, giving Eq.\eqref{Psuccess} as
\begin{equation}
P_{succ}=\frac{1}{2}\left[1+ \left(\eta_0-\eta_1\right) \, \left|1-2r\right| \right].
\end{equation}
The optimal input will be characterized by values $r=0$ or $r=1$.


\subsection{Amplitude damping channel}

The qubit amplitude damping channel is described by the map  \cite{MWbook}
\be\label{calN}
\psi\mapsto{\cal A}_\eta(\psi)=(|0\rangle\langle 0|+\cos\eta |1\rangle\langle 1|) \psi 
(|0\rangle\langle 0|+\cos\eta |1\rangle\langle 1|)
+\sin^2\eta  |0\rangle\langle 1|\psi |1\rangle\langle 0|,
\ee
with $\eta\in[0,\frac{\pi}{2}]$.

For the input state, w.l.o.g. we can confine our attention to the $x-z$ plane
(because of the symmetric action of ${\cal A}$ 
with respect to rotations around the $z$ axis in the Bloch sphere). Hence, we can choose $\varphi=0$. 

This yields for Eq.\eqref{Psuccess} 
\begin{align}\label{Psuccx}
P_{succ}=\frac{1}{2}\left[1+(\cos\eta_1-\cos\eta_0)\sqrt{
r^2(\cos\eta_0+\cos\eta_1)^2-r^2+r}\right].
\end{align}
Maximizing \eqref{Psuccx} over $r$, while taking into account that $0\leq r\leq 1$, we finally get 
\begin{subequations}\label{Psuccnofb}
\begin{align}
P_{succ}&=\frac{1}{4}\left[2+ \frac{ \cos \eta_1-\cos\eta_0}{\sqrt{1-(\cos\eta_0+\cos\eta_1)^2}}\right],
&\text{for} \quad  (\cos\eta_0+\cos\eta_1) < \frac{1}{\sqrt 2}, \\
P_{succ}&=\frac{1}{2}\left[ \sin^2\eta_0+\cos^2\eta_1\right],
&\text{for} \quad  (\cos\eta_0+\cos\eta_1) \geq \frac{1}{\sqrt 2}.
\end{align}
\end{subequations}
The optimal value of $r$ results in
\begin{subequations}\label{optx1shot}
\begin{align}
r&=\frac{1}{2(1-\gamma^2)},
\hspace{-1.5cm}&\text{for} \quad (\cos\eta_0+\cos\eta_1)<\frac{1}{\sqrt{2}}, \\
r&=1,
\hspace{-1.5cm}&\text{for} \quad (\cos\eta_0+\cos\eta_1)\geq\frac{1}{\sqrt{2}}.
\end{align}
\end{subequations}

}


\section{Multi-shot discrimination}

We are now going to consider the multi-shot discrimination
{ assuming local input states. First, we describe the global measurement strategy that will serve as a benchmark (see Fig.\ref{fig1} right). Then, we assume
local measurements with classical information feedforward and no quantum memory (see Fig.\ref{fig1} left).} At each step, we update the observable $\Delta$, defined as in \eqref{eq:Delta}, by updating $p_0$, the weight of $\rho_0$, consequently the weight of $\rho_1$, and apply the Helstrom measurement as we do in the single-shot. 
Actually, we shall consider the measurement adjusted at each step based on the previous outcomes. 
Two strategies can be figured out: one in which all previous outcomes are accounted for (Bayesian strategy), and the other in which only the immediately preceding outcome is used (Markovian strategy). 
\begin{figure}[H]
\centering
\begin{subfigure}[t]{0.45\textwidth}
\begin{quantikz}[row sep=1cm,column sep=0.5cm,wire types={q,q,n,q,q,q}]
\lstick{${\psi(r_1)}$}& \gate{ ? } &\push{\rho_{?}(r_1)} &\meter{} \\ 
\lstick{${\psi(r_2)}$}& \gate{ ? } &\push{\rho_{?}(r_2)}& \meter{}  \\ 
{\vdots} & {\vdots}  &  {\vdots}&  {\vdots} \\
\lstick{${\psi(r_{n+1})}$}& \gate{ ? } &\push{\rho_{?}(r_{n+1})}& \meter{}  \\ 
\arrow[from=1-4, to=2-1, style={dashed, arrows=->}]
\arrow[from=1-4, to=2-4, style={dashed, arrows=->}]
\arrow[from=2-4, to=3-1, style={dashed, arrows=->}]
\arrow[from=2-4, to=3-4, style={dashed, arrows=->}]
\arrow[from=3-4, to=4-1, style={dashed, arrows=->}]
\arrow[from=3-4, to=4-4, style={dashed, arrows=->}]
\end{quantikz}   
\end{subfigure}
\begin{subfigure}[t]{0.45\textwidth}
\begin{quantikz}[row sep=1cm,column sep=0.5cm,wire types={q,q,n,q,q,q}]
\lstick{${\psi(r_1)}$}& \gate{? } &\push{\rho_{?}(r_1)} & \meter[4]{} \\ 
\lstick{${\psi(r_2)}$}& \gate{ ? } &\push{\rho_{?}(r_2)}&\qw  \\ 
{\vdots} & {\vdots}  &  {\vdots} \\
\lstick{${\psi(r_{n+1})}$}& \gate{ ? } &\push{\rho_{?}(r_{n+1})}&\qw  \\ 
\end{quantikz}    
\end{subfigure}
\caption{ Left: Schematic representation of channel discrimination through a local adaptive strategy. $\psi$ (resp. $\rho$) denotes the input (resp. output) state at each step. The symbol "?" stands for the quantum channel picked up from a binary ensemble. Dashed (resp. solid) lines refer to the flow of classical (resp. quantum) information. Right: Schematic representation of channel discrimination with global measurement.}
\label{fig1}
\end{figure}

{
\subsection{Global strategy}

We begin by considering $n+1$ parallel uses of the channel and assuming that
at each use there is no prior knowledge about which of the two channels is being applied (see Fig.\ref{fig1} right).
This implies having at each output $\rho_0$ or $\rho_1$ 
with probability $\frac{1}{2}$.
Then, taking into account that the $i$th output depends on the input parameter $r_i$,
we have to consider the following operator, generalizing Eq.\eqref{eq:Delta}, 
\begin{equation}
\boldsymbol{\Delta}:=\frac{1}{2}\left[\otimes_{i=1}^{n+1}\rho_0(r_i)- \otimes_{i=1}^{n+1}\rho_1(r_i)\right].
\end{equation}
Here, the bold symbol denotes operators acting on multiple (actually $n+1$) qubit space.
From the eigenvalues $\lambda_i$ and orthogonalized eigenvectors $|\lambda_i\rangle$ of $\boldsymbol{\Delta}$ we construct the POVM elements
\begin{subequations}
\begin{align}
\boldsymbol{\Pi}_0&:=\sum_{i: \lambda_i\geq 0}|\lambda_i\rangle\langle\lambda_i|,\\
\boldsymbol{\Pi}_1&:=\boldsymbol{I}-\boldsymbol{\Pi}_0.
\end{align}
\end{subequations} 
Then, the success probability is given by
\begin{equation}\label{Psglobal}
P_{succ}^{(n+1)}=\frac{1}{2}\left[{\rm Tr}\left(\otimes_{i=1}^{n+1}\rho_0(r_i)\boldsymbol{\Pi}_0\right)
+{\rm Tr}\left(\otimes_{i=1}^{n+1}\rho_1(r_i)\boldsymbol{\Pi}_1\right)\right].
\end{equation}
If all the eigenvalues $\lambda_i$ have the same sign, we will use the projector $\boldsymbol{I}$ or $\boldsymbol{0}$ in place of $\boldsymbol{\Pi}_0$ as discussed in the one-shot case.

}

\subsection{Bayesian strategy}

{  Also in this case, we consider $n+1$ parallel uses of the channel. However, we assign equal prior probability $\frac{1}{2}$ to the two channels only at the first use. For each subsequent use, the probabilities assigned to $\rho_0$ and $\rho_1$ will depend on all previous outcomes (see Fig.\ref{fig1} left).}

Let us denote by $x^n$ the sequence $x_1x_2\ldots x_n\in\{0,1\}^n$ of measurement outcomes.
At the step $n+1$ (with $n>0$), we can write the success probability as
\begin{equation}\label{PsB}
P_{succ}^{(n+1)}=\sum_{x^n\in\{0,1\}^n} P_{\rho_0}(x^n,0)
+\sum_{x^n\in\{0,1\}^n} P_{\rho_1}(x^n,1),
\end{equation}
where 
\begin{align}
P_\rho(x^n,0)&={\rm Tr}\left(\rho \Pi_{x_1}\right){\rm Tr}\left(\rho \Pi^{x_1}_{x_2}\right)\ldots
{\rm Tr}\left(\rho \Pi_{x_n}^{x_1\ldots x_{n-1}}\right){\rm Tr}\left(\rho \Pi_{x_{n+1}=0}^{x_1\ldots x_{n-1}x_n}\right),\\
P_\rho(x^n,1)&={\rm Tr}\left(\rho \Pi_{x_1}\right){\rm Tr}\left(\rho \Pi^{x_1}_{x_2}\right)\ldots
{\rm Tr}\left(\rho \Pi_{x_n}^{x_1\ldots x_{n-1}}\right){\rm Tr}\left(\rho \Pi_{x_{n+1}=1}^{x_1\ldots x_{n-1}x_n}\right).
\end{align}
The maximization of $P_{succ}^{(n+1)}$ can be split into $2^n$ maximization problems
for finding the POVMs $\{\Pi_{0}^{x^n},\Pi_{1}^{x^n}\}$ (notice that the label of each POVM is $x^n\in\{0,1\}^n$).

Each of these problems, specified by the sequence $x^n$, is solved by means of the eigenvectors of the operator
\begin{equation}\label{opMBayes}
\Delta^{(n+1)}:=\frac{P_{\rho_0}(x^n)}{P_{\rho_0}(x^n)+P_{\rho_1}(x^n)} \rho_0
- \frac{P_{\rho_1}(x^n)}{P_{\rho_0}(x^n)+P_{\rho_1}(x^n)} \rho_1,
\end{equation}
with
\begin{equation}
P_\rho(x^n)={\rm Tr}\left(\rho \Pi_{x_1}\right){\rm Tr}\left(\rho \Pi^{x_1}_{x_2}\right)\ldots
{\rm Tr}\left(\rho \Pi_{x_n}^{x_1\ldots x_{n-1}}\right).
\end{equation}
The (normalized) eigenvectors of Eq.\eqref{opMBayes} provide the POVM elements $\{ \Pi_{0}^{x_1\ldots x_n},
 \Pi_{1}^{x_1\ldots x_n}\}$.

Thus, the success probability can be computed iteratively starting from the first step (corresponding to $n=0$) where the POVM $\{\Pi_{0},\Pi_{1}\}$ is determined by the eigenvectors of $\Delta^{(1)}=\frac{1}{2}\rho_0-\frac{1}{2}\rho_1$.


\subsection{Markovian strategy}

{  As in the Bayesian case, we consider $n+1$ parallel uses of the channel and assign equal probability $\frac{1}{2}$ to the appearance of either channel only at the first use. For each subsequent use, the probability assigned to
$\rho_0$ and consequently to $\rho_1$, depends on the outcome of the immediately preceding use (see Fig.\ref{fig1} left).}

At the step $n+1$ (with $n>0$), we can write the success probability as Eq.\eqref{PsB}
\begin{equation}\label{PsM}
P_{succ}^{(n+1)}=\sum_{x^n\in\{0,1\}^n} P_{\rho_0}(x^n,0)
+\sum_{x^n\in\{0,1\}^n} P_{\rho_1}(x^n,1).
\end{equation}
However, this time we have
\begin{align}
P_\rho(x^n,0)&={\rm Tr}\left(\rho \Pi_{x_1}\right){\rm Tr}\left(\rho \Pi^{x_1}_{x_2}\right)\ldots
{\rm Tr}\left(\rho \Pi_{x_n}^{x_{n-1}}\right){\rm Tr}\left(\rho \Pi_{x_{n+1}=0}^{x_n}\right),\\
P_\rho(x^n,1)&={\rm Tr}\left(\rho \Pi_{x_1}\right){\rm Tr}\left(\rho \Pi^{x_1}_{x_2}\right)\ldots
{\rm Tr}\left(\rho \Pi_{x_n}^{x_{n-1}}\right){\rm Tr}\left(\rho \Pi_{x_{n+1}=1}^{x_n}\right).
\end{align}
The maximization of $P_{succ}^{(n+1)}$ can be split into $2$ maximization problems
for finding the POVMs $\{\Pi_{0}^{x_n},\Pi_{1}^{x_n}\}$ (notice that the label of each POVM is $x_n\in\{0,1\}$).

Each of these problems, specified by the bit value $x_n$, is solved by the eigenvectors of the operator 
\begin{align}\label{opMMarkov}
\Delta^{(n+1)}&:=
\frac{\sum_{x^{n-1}}P_{\rho_0}(x^{n-1},x_n)}{
\sum_{x^{n-1}}P_{\rho_0}(x^{n-1},x_n)+\sum_{x^{n-1}}P_{\rho_1}(x^{n-1},x_n)}
\rho_0 \notag\\
&- \frac{\sum_{x^{n-1}}P_{\rho_1}(x^{n-1},x_n)}{
\sum_{x^{n-1}}P_{\rho_0}(x^{n-1},x_n)+\sum_{x^{n-1}}P_{\rho_1}(x^{n-1},x_n)}
\rho_1.
\end{align}
The (normalized) eigenvectors of Eq.\eqref{opMMarkov} provides the POVM elements $\{ \Pi_{0}^{x_n},
 \Pi_{1}^{x_n}\}$.

Thus, the success probability can be computed iteratively starting from the first step (corresponding to $n=0$) where the POVM $\{\Pi_{0},\Pi_{1}\}$ is determined by the eigenvectors of $\Delta^{(1)}=\frac{1}{2}\rho_0-\frac{1}{2}\rho_1$.


\bigskip


{
It is clear that $P_{succ}^{(n+1)}$ in all strategies (Eqs.\eqref{Psglobal}, \eqref{PsB} and \eqref{PsM}) depends on the input state at each step, given that 
$\rho_0$ and $\rho_1$ correspond to the action of two different channels on $\psi(r)$. 
Thus, the probability $P_{succ}^{(n+1)}$ should finally be optimized over a multi-parameter $r^{n+1}\equiv(r_1, r_2, \ldots, r_{n+1})$.

Notice that, due to this input optimization, the Markovian and Bayesian strategies here described include the case where also the input is chosen according to the information gathered from previous step(s) (see diagonal dashed lines in Fig.\ref{fig1} left).

}


\section{Results}

To evaluate the performance of Bayesian and Markovian strategies described above, we start considering the first non-trivial case of three-shot discrimination, corresponding to $n=2$ (for two-shot discrimination the Bayesian strategy coincides with the Markovian one).

{ 
Then, for specific points in the parameters' region we compare the performance of
global, Bayesian and Markovian strategies in terms of the number of shots \footnote{Calculations beyond $n=8$ (nine shots) are computationally not affordable due to the rapid increase of the dimensions of the space of input states. }.

Below, we will distinguish the results for the three types of channels.}

{
\subsection{Depolarizing channel}

In Fig.\ref{fig:depo}, it is shown the density plot of the difference between Eqs.\eqref{PsB} and \eqref{PsM} vs $\eta_0,\eta_1$ when $n=2$. 
We can see that the difference between Bayesian and Markovian strategies appears in the top right corner.

\begin{figure}[H]
  \centering
    \includegraphics[width=8cm]{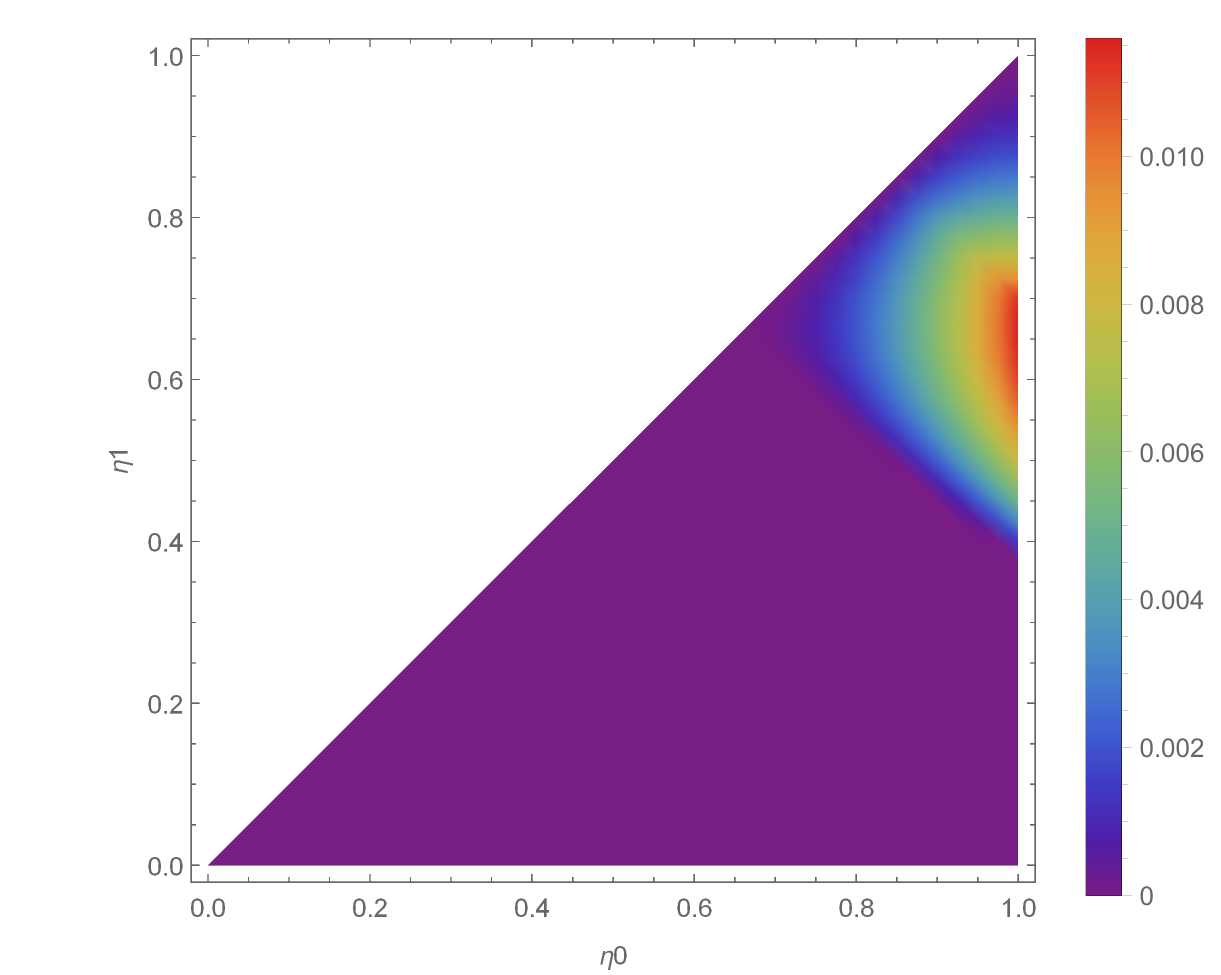}
    \caption{ Difference between $P_{succ}^{(3)}$ for Bayesian strategy and $P_{succ}^{(3)}$ for Markovian strategy vs $\eta_0,\eta_1$ for depolarizing channels. }
    \label{fig:depo}
\end{figure}

In Fig.\ref{fig:2} the success probabilities obtained with global, Bayesian, and Markovian strategies are shown versus $n$. 
We may notice that no matter which point in the plane $\eta_0,\eta_1$ we take, Bayesian and Markovian strategies perform almost the same.

Notice that, for depolarizing channels, we do not need to maximize over the input states, because the success probability of the various strategies does not depend on them. Thus, we can always use the same input -- e.g. $|0\rangle$.

\begin{figure}[H]
\begin{subfigure}[b]{0.45\textwidth}
    \includegraphics[width=\linewidth]{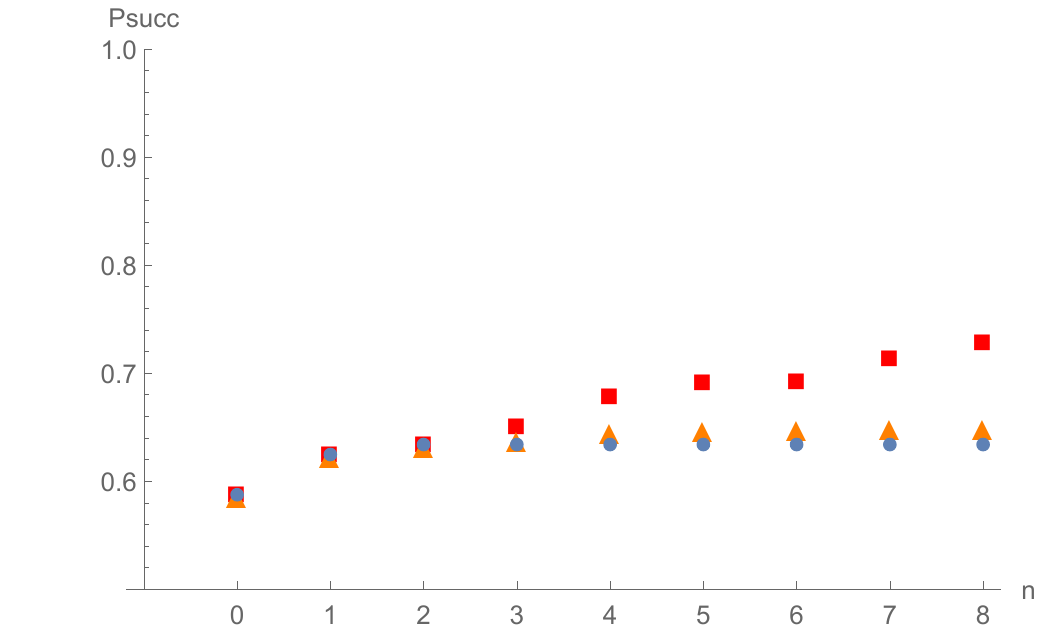}
\caption{ $\eta_0=0.75,\eta_1=0.4$}
    \label{fig:2a}
\end{subfigure}
\begin{subfigure}[b]{0.45\textwidth}
    \includegraphics[width=\linewidth]{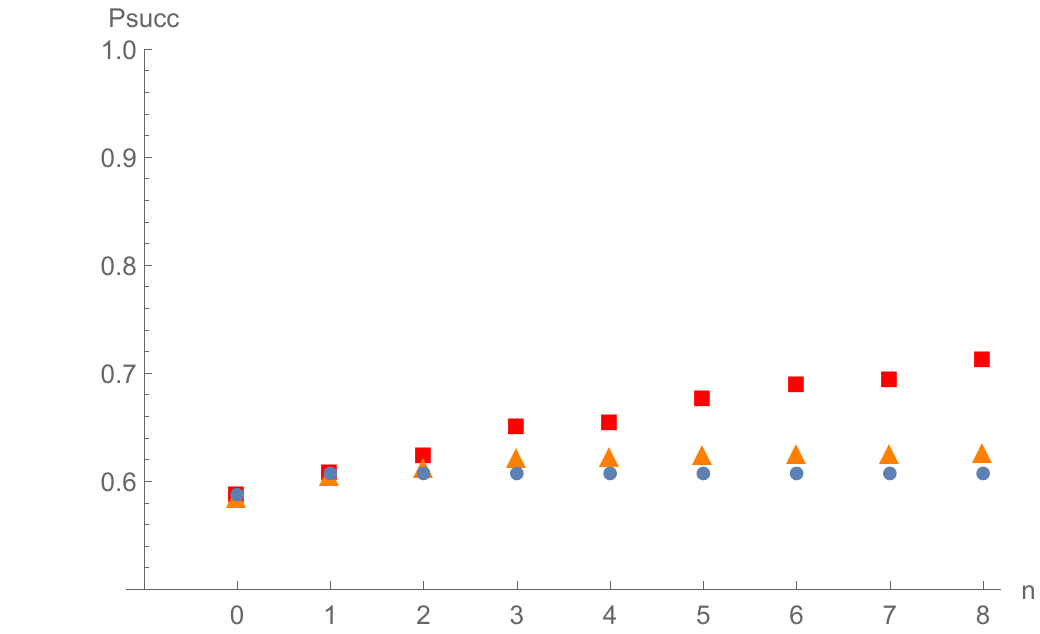}
\caption{ $\eta_0=0.95,\eta_1=0.6$}
    \label{fig:2b}
    \end{subfigure}
    \caption{ Success probability $P_{succ}^{(n)}$ vs $n$ computed in different points (a), (b) of the $\eta_0,\eta_1$ plane for depolarizing channels. Red squares, Orange triangles, Blue dots correspond to global, Bayesian and Markovian strategies respectively.}
    \label{fig:2}
\end{figure}
}

{
\subsection{Bit-flip channel}

In Fig.\ref{fig:bit} the density plot of the difference between Eqs.\eqref{PsB} and \eqref{PsM} for $n=2$ vs $\eta_0,\eta_1$ optimized over parameters $r_1, r_2, r_3$ is shown. 
We can see that the difference between Bayesian and Markovian strategies appears only around the line $\eta_1=\pi/2-\eta_0$ (anti-diagonal), and it is almost one order of magnitude larger than that for the depolarizing channel.

\begin{figure}[H]
  \centering
    \includegraphics[width=8cm]{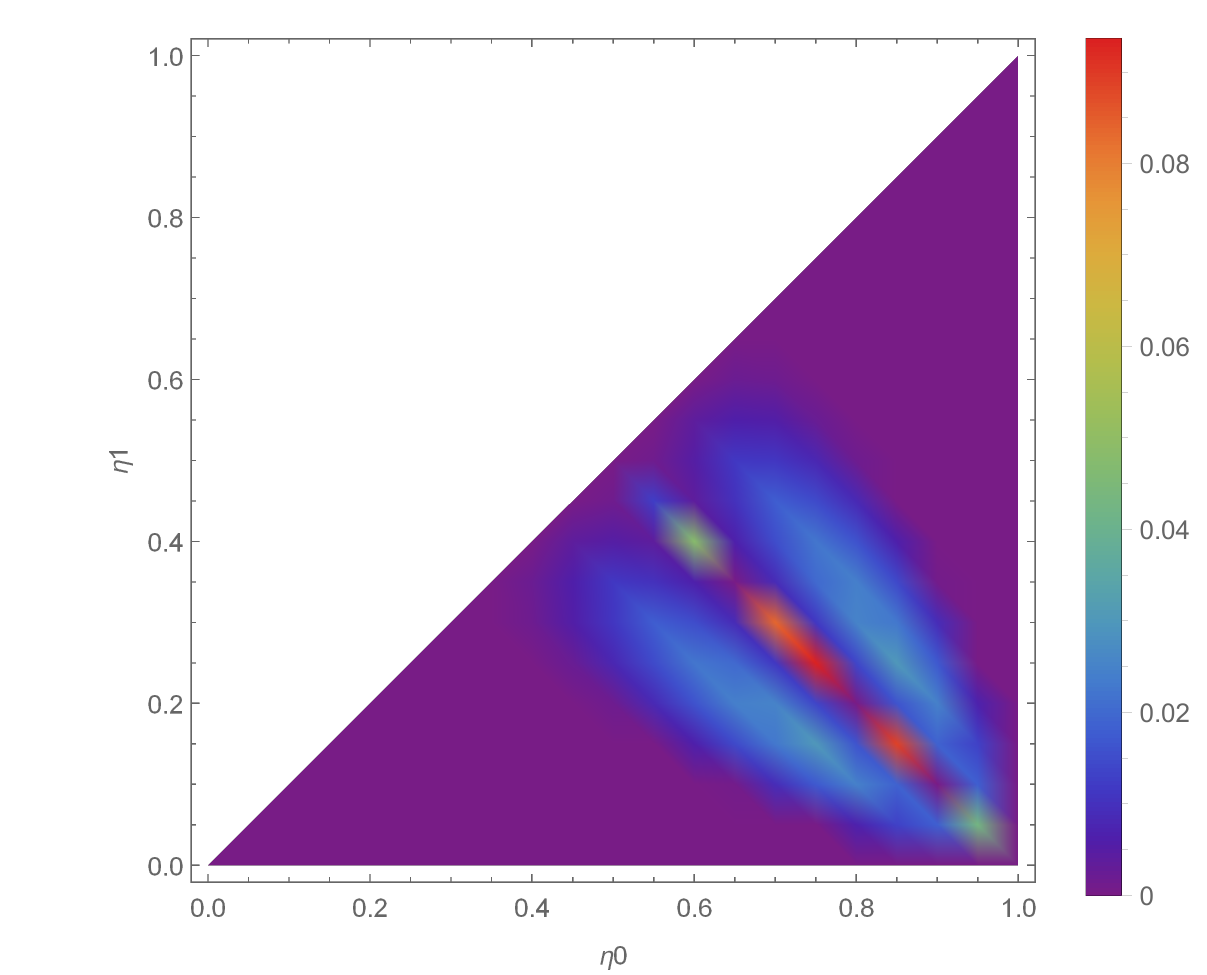}
    \caption{ Difference between $P_{succ}^{(3)}$ for Bayesian strategy and $P_{succ}^{(3)}$ for Markovian strategy vs $\eta_0,\eta_1$ for bit-flip channels. }
    \label{fig:bit}
\end{figure}

Fig.\ref{fig:4} presents the success probabilities obtained with global, Bayesian, and Markovian strategies versus $n$. 
Fig.\ref{fig:4a} shows an appreciable difference between Bayesian and Markovian strategies compared to Fig.\ref{fig:4b}, due to point (a) being located along the anti-diagonal of the parameters' region.   

For point (b), away from anti-diagonal, the Bayesian and Markovian strategies yield identical results up to $n=6$, and both coincide with the global strategy up to $n=4$. This demonstrates that the proposed strategies are more effective w.r.t. the the case of depolarizing channels (where coincidence with global strategy holds up until $n=2$).

For all success probabilities a numerical maximization over parameters $r_i$ is done. In fact, the optimal input states in multi-shot scenarios are not necessarily constrained to have parameters $r_i$ equal to 0 or 1 like in the one-shot case. 

\begin{figure}[H]
\begin{subfigure}[b]{0.45\textwidth}
    \includegraphics[width=\linewidth]{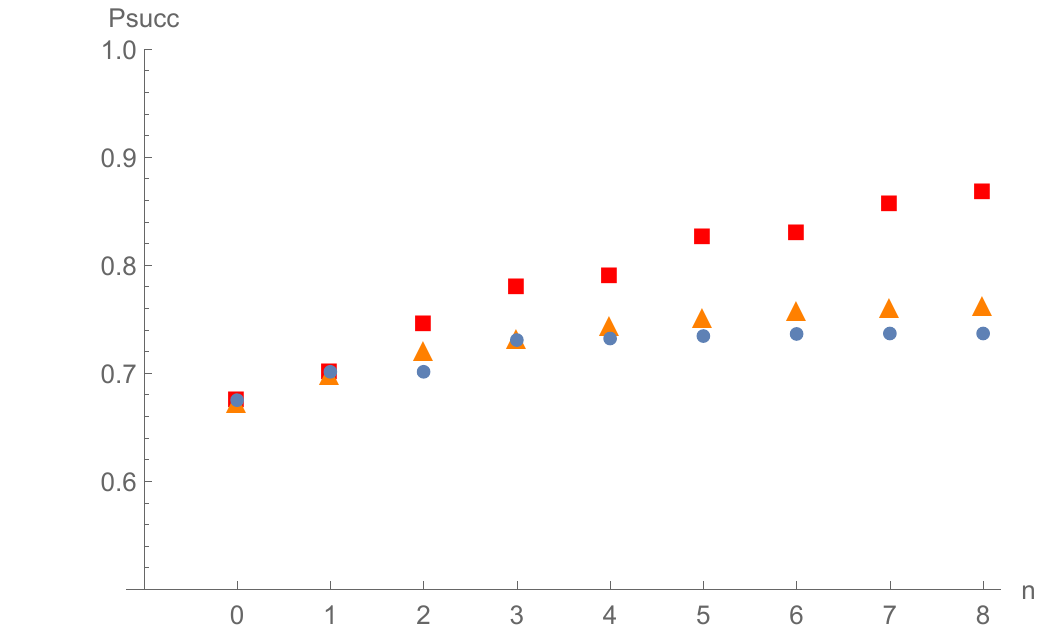}
\caption{ $\eta_0=0.75,\eta_1=0.4$}
    \label{fig:4a}
\end{subfigure}
\begin{subfigure}[b]{0.45\textwidth}
    \includegraphics[width=\linewidth]{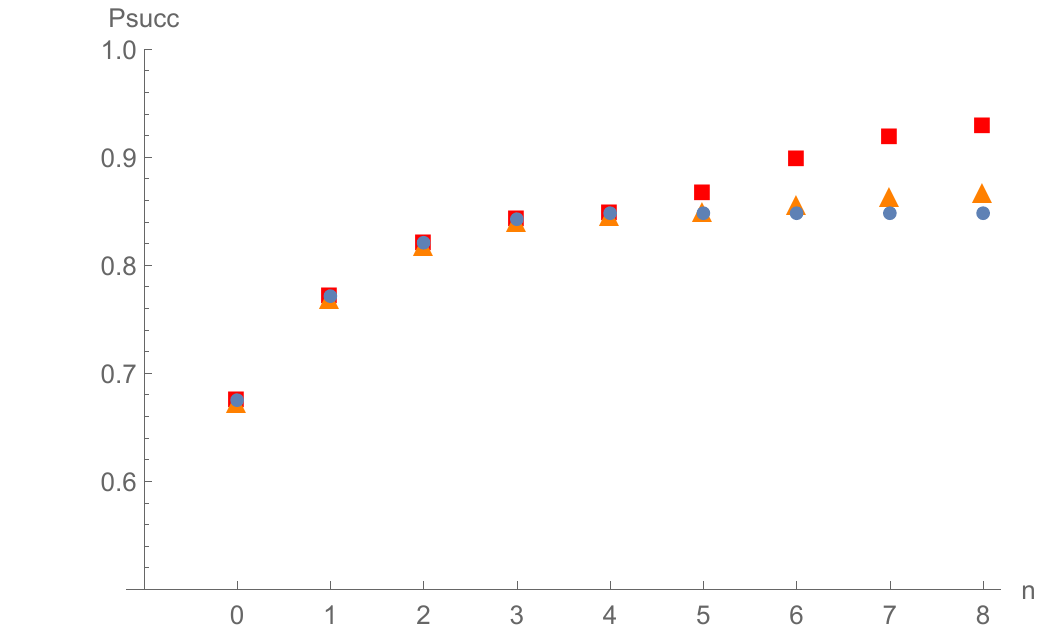}
\caption{ $\eta_0=0.95,\eta_1=0.6$}
    \label{fig:4b}
    \end{subfigure}
    \caption{ Success probability $P_{succ}^{(n)}$ vs $n$ computed in different points (a), (b) of the $\eta_0,\eta_1$ plane for bit-flip channels. Red squares, Orange triangles, Blue dots correspond to global, Bayesian and Markovian strategies respectively.}
    \label{fig:4}
\end{figure}

}

{
\subsection{Amplitude damping channel}
}

\begin{figure}[H]
  \centering
    \includegraphics[width=8cm]{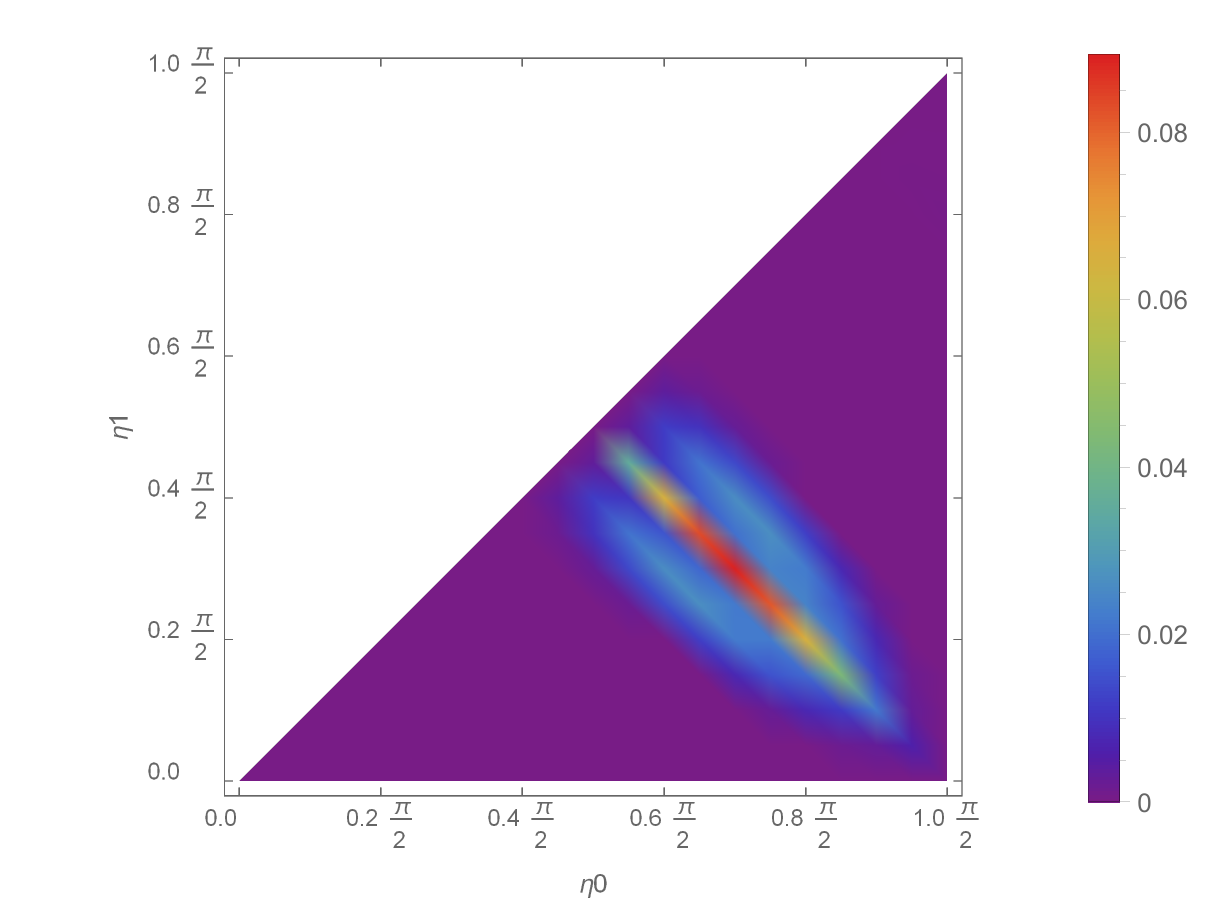}
    \caption{Difference between $P_{succ}^{(3)}$ for Bayesian strategy and $P_{succ}^{(3)}$ for Markovian strategy vs $\eta_0,\eta_1$, for amplitude damping channels.
   }
    \label{fig:m2}
\end{figure}

In Fig.\ref{fig:m2} it is shown the { density} plot of the difference between Eqs.\eqref{PsB} and \eqref{PsM} for $n=2$ vs $\eta_0,\eta_1$ optimized over parameters $r_1,r_2,r_3$. We can see that the difference between Bayesian and Markovian strategies appears only around the line $\eta_1=\pi/2-\eta_0$ (anti-diagonal). { The behavior is similar to that of bit-flip channels even in terms of magnitude.}

\begin{figure}[ht]
\begin{subfigure}[b]{0.45\textwidth}
    \includegraphics[width=\linewidth]{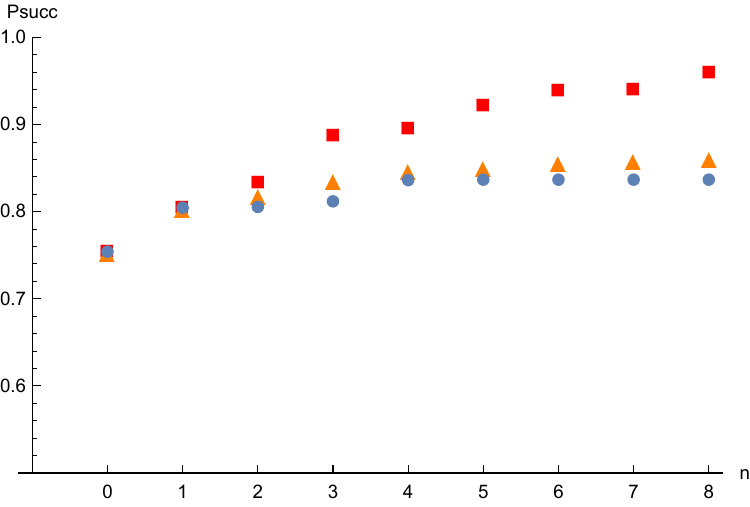}
\caption{ $\eta_0=0.75\frac{\pi}{2},\eta_1=0.4\frac{\pi}{2}$}
    \label{fig:3b}
\end{subfigure}
\begin{subfigure}[b]{0.45\textwidth}
    \includegraphics[width=\linewidth]{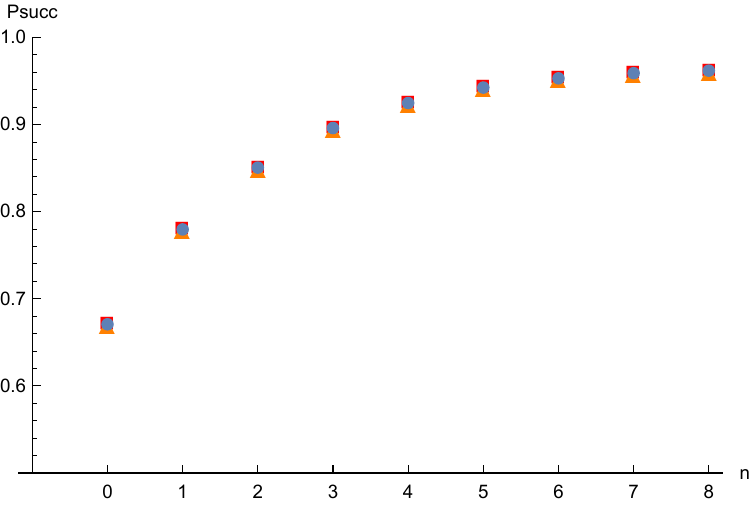}
\caption{ $\eta_0=0.95\frac{\pi}{2},\eta_1=0.6\frac{\pi}{2}$}
    \label{fig:3c}
    \end{subfigure}
    \caption{ Success probability $P_{succ}^{(n)}$ vs $n$ computed in different points (a), (b) of the $\eta_0,\eta_1$ plane for amplitude damping channels. Red squares, Orange triangles, Blue dots correspond to global, Bayesian and Markovian strategies respectively.}
    \label{fig:6}
\end{figure}

{ In Fig.\ref{fig:6} shows that even at point (a)—which corresponds to the maximum value in Fig.\ref{fig:m2}—the difference between the Bayesian and Markovian strategies remains relatively small over $n$.

As we move away from point (a) (in the anti-diagonal), the differences among the strategies diminish rapidly. Actually, at point (b), the success probability of the Markovian strategy not only closely matches that of the Bayesian strategy, but also aligns with the global strategy.

For all success probabilities a numerical maximization over parameters $r_i$ is done. In fact, although point (a) lies within the parameter region where the optimal input for the single-shot is $\ket{1}$, the optimal input state for subsequent shots may vary. This is due to the fact that the boundary of Eq.\eqref{optx1shot} is influenced by the prior probabilities of the states.

}

\section{Conclusion}

In conclusion, we have investigated the multi-shot discrimination between { two qubit channels-- be either depolarizing, or bit-flip or amplitude damping--using separable inputs, and without side entanglement, employing Helstrom measurement at each step along with
classical information feedforward.}
We compared the performance of Bayesian and Markovian strategies with local inputs determined through the maximization of the final success probability. Our findings show that the former is only slightly advantageous and for a limited parameters region { (that on the line $\eta_1=\pi/2-\eta_0$ for bit-flip channels and amplitude damping channels).}

{ The fact that the success probability of the devised adaptive strategies does not consistently reach that of the global strategy
suggests that the chosen adaptive strategy is not always optimal. While there are instances--particularly for small $n$--where the success probabilities coincide, 
for large $n$ we expect the performance of optimal global and optimal local adaptive strategies to converge\cite{SHW22}. This discrepancy likely arises because
optimizing each measurement step individually does not lead to overall optimality across $n+1$-steps. In other words, achieving the highest final success probability may require "sacrifices" at earlier steps.
Interestingly, the possibility of attaining the same performance of the global strategy appears to be a peculiarity of the amplitude damping channels (see Fig.\ref{fig:3c}), which may be attributed to their non-unital nature.}

For our Bayesian strategy we pursued an updating process using forward optimization; however, in some cases, backward optimization may yield better performance \cite{Hetal11}. A comparative analysis of the two approaches is left for future work. Nonetheless, the Markovian strategy proves to be highly effective and warrants further study for the discrimination of other types of quantum channels { beyond the qubit space. It could be also employed in the emerging direction of determining the minimum number of shots to achieve a specific error rate requirement \cite{Tian23}.

}


\subsection*{Acknowledgments}
The authors acknowledge financial support from the ``PNRR MUR project PE0000023-NQSTI''.

\end{document}